\documentclass[aps,pra,twocolumn,superscriptaddress,floatfix,longbibliography]{revtex4}
\usepackage{amsfonts}
\usepackage{amssymb}
\usepackage{graphicx}
\usepackage{dcolumn}
\usepackage{bm}
\usepackage{amsmath}
\usepackage[colorlinks,linkcolor=magenta,citecolor=blue,urlcolor=blue]{hyperref}
\usepackage{changes}
\usepackage{xcolor}
\usepackage{ulem} 
\begin{document}
	
\preprint{This line only printed with preprint option}
	
\title{Multiple reentrant topological windows induced by generalized Bernoulli disorder}
	
\author{Ruijiang Ji}
\affiliation {Institute of Theoretical Physics and State Key Laboratory of Quantum Optics Technologies and Devices, Shanxi University, Taiyuan 030006, China}
	
\author{Yunbo Zhang}
\email[Contact author: ]{ybzhang@zstu.edu.cn}
\affiliation {Zhejiang Key Laboratory of Quantum State Control and Optical Field Manipulation, Department of Physics, Zhejiang Sci-Tech University, Hangzhou 310018, China}
	
\author{Shu Chen}
\email[Contact author: ]{schen@iphy.ac.cn}
\affiliation {Beijing National Laboratory for Condensed Matter Physics, Institute of Physics, Chinese Academy of Sciences, Beijing 100190, China}
\affiliation {School of Physical Sciences, University of Chinese Academy of Sciences, Beijing 100049, China}
	
\author{Zhihao Xu}
\email[Contact author: ]{xuzhihao@sxu.edu.cn}
\affiliation {Institute of Theoretical Physics and State Key Laboratory of Quantum Optics Technologies and Devices, Shanxi University, Taiyuan 030006, China}
\affiliation {Collaborative Innovation Center of Extreme Optics, Shanxi University, Taiyuan 030006, China}
	
\date{\today}
\begin{abstract}
We investigate reentrant topological transitions in a one-dimensional Su-Schrieffer-Heeger chain with generalized Bernoulli disorder in the intradimer hopping amplitudes. Owing to its independently tunable values and probabilities, the multivalued disorder distribution provides a direct way to control the topological phase diagram. We show that increasing the disorder strength can split the nontrivial regime into multiple disconnected topological windows, whose number, widths, and locations are determined by the distribution parameters. The phase boundaries are derived analytically from the zero-mode inverse localization length and are governed by a weighted geometric mean of the disordered hopping amplitudes, in agreement with numerical results from the reflection-matrix topological quantum number and the real-space winding number. We also show that the mean chiral displacement dynamically identifies these reentrant windows. These results demonstrate how multivalued random disorder can organize and tune reentrant topological behavior in one-dimensional chiral lattices.
\end{abstract}

\maketitle

\section{Introduction}
%{\it Introduction.} 
Topological insulators and related topological phases have attracted sustained interest because of their robust boundary states and unconventional transport properties~\cite{HasanMZ2010,QiXL2011,LuL2014,OzawaT2019,KrausYE2012,VerbinM2015,StutzerS2018,DengJ2022,St-JeanP2017,PartoM2018,ZhaoH2018,DaiT2024,WangW2025,ChengX2016}. In disordered systems, the interplay between topology and localization can generate phenomena absent in clean lattices. A representative example is the topological Anderson insulator (TAI), in which disorder drives a topologically trivial system into a nontrivial one~\cite{AndersonPW1958,LiJ2009,JiangH2009,GrothCW2009,GuoHM2010,SongJ2012,TitumP2015,TangLZ2020,ZhangW2021,PengT2021,CuiX2022,LinQ2022,ChengX2023,RenM2024,SobrosaN2024,AssuncaoBD2024,ZuoZW2024}. Such disorder-induced topological behavior~\cite{VeluryS2021,NomuraK2011,EversF2008,LiHL2021,PordanE2010,ChengX2026,KrishtopenkoSS2022,HuYS2021,WuHB2021,OngZY2016,SongJ2014,GirschikA2013,LiCA2020,ChenA2024,WangJH2021,WangC2022,GrindallC2025,RoyK2024,RoyS2023} has been explored in a variety of platforms, including cold atoms~\cite{MeierEJ2018}, photonic lattices~\cite{StutzerS2018,LiuGG2020,LinQ2022}, and electric circuits~\cite{ZhangW2021}. Beyond conventional Anderson-type randomness, quasiperiodic and other structured disorders can also induce topological transitions and rich localization behavior~\cite{LonghiS2020,ZhangGQ2021,TangLZ2022,LuZ2022,LiX2024,WangXM2025,SinhaA2025,LiuSN2022,GhoshAK2024,JiR2025}. Understanding how the structure of disorder reshapes topological phase diagrams remains an active topic.

Recent studies have revealed that one-dimensional systems with quasiperiodic, aperiodic, or other structured modulations can exhibit reentrant topological behavior, where the system enters, leaves, and re-enters a topological phase as a control parameter is varied~\cite{RoyK2024,RoyS2023,RoyS2021,RoyS2022,ZuoZW2022,LuZ2025,BanerjeeS2026,ZhangXY2026}. Such behavior produces disconnected topological regions in parameter space and is often accompanied by rich localization properties. Existing studies have mainly considered continuous random disorder or deterministic quasiperiodic and aperiodic modulations. By contrast, much less attention has been paid to multivalued discrete random disorder, although it offers a simple way to tune not only the disorder strength but also the statistical composition of the disordered couplings.

A generalized Bernoulli distribution provides a particularly useful setting for this purpose. It consists of several discrete disorder values with independently tunable probabilities, and therefore allows one to separate the effects of disorder amplitudes from those of statistical weights. This raises a natural question: how do the values and probabilities of a discrete random distribution determine the number, widths, and positions of reentrant topological windows? For chiral one-dimensional systems, this question is especially tractable because the zero-mode wave function is governed by products of hopping amplitudes. As a result, the relevant quantity controlling the topological transition is not the arithmetic average of the disorder, but a weighted geometric mean of the effective hopping amplitudes.

In this work, we address this question in a one-dimensional Su-Schrieffer-Heeger (SSH) chain with generalized Bernoulli-type disorder in the intradimer hopping amplitudes. We show that a multivalued random distribution can split the topological regime into multiple disconnected windows as the disorder strength is varied. The phase boundaries are derived analytically from the inverse localization length of the zero modes and agree well with numerical calculations. We further demonstrate that the number of reentrant topological windows can be increased by using disorder distributions with more components, while their widths and center positions can be tuned by the probabilities and relative amplitudes of these components. Finally, we show that the mean chiral displacement provides a dynamical signature of these disorder-induced topological transitions. Our results therefore identify generalized Bernoulli disorder as a simple and analytically controllable mechanism for organizing reentrant topology in one-dimensional chiral lattices.

\section{Model and method}
%{\it Model and method.} 
We consider a one-dimensional SSH chain with uniform interdimer hopping $t_2$ and generalized Bernoulli-type disorder introduced in the intradimer hopping $t_1$. The Hamiltonian is
\begin{equation}\label{Eq.1}
\hat{H}=-\sum_{i}\left[(t_1-\xi_i)\hat{c}_{i,A}^{\dagger}\hat{c}_{i,B}+t_2\hat{c}_{i,B}^{\dagger}\hat{c}_{i+1,A}+\textrm{H.c.}\right],
\end{equation}
where $\hat{c}_{i,\sigma}$ ($\hat{c}_{i,\sigma}^{\dagger}$) annihilates (creates) a particle on sublattice $\sigma=A,B$ in the $i$th unit cell. The random variable $\xi_i$ is drawn independently from a discrete distribution with $M$ possible values, denoted by $\xi^{(1)},\xi^{(2)},\ldots,\xi^{(M)}$, which occur with probabilities $p_1,p_2,\ldots,p_M$, respectively, satisfying $\sum_j p_j=1$. Equivalently, its probability distribution is written as $P(\xi_i)=\sum_{j=1}^{M} p_j \delta(\xi_i-\xi^{(j)})$. Throughout this work, we set $t_2=1$ as the energy unit.

Since translational symmetry is broken by disorder, we characterize the topology using the zero-energy reflection-matrix topological quantum number for one-dimensional chiral-symmetric systems~\cite{FulgaIC2011,Zhang2016}, which can be written as
\begin{equation}\label{Eq.2}
Q=\frac{1}{2}\left(1-\mathrm{sgn}\left[\prod_{i}\left(-t_1+\xi_i\right)^2-\left[t_2\right]^{2N}\right]\right).
\end{equation}
This quantity remains well defined in the absence of translational symmetry and diagnoses the presence or absence of robust zero-energy end modes. In particular, $Q=1$ corresponds to a topologically nontrivial phase with zero-energy boundary modes, while $Q=0$ corresponds to a trivial phase (see Appendix A for details). To reduce sample-to-sample fluctuations, we further define the disorder-averaged topological quantum number as $\overline{Q}=N_c^{-1}\sum_{c=1}^{N_c}Q_c$, where $Q_c$ is the value of $Q$ for the $c$-th disorder realization and $N_c$ is the total number of disorder realizations. We have also verified the same phase boundaries using the disorder-averaged real-space winding number~\cite{RoyK2024,RoyS2023} (see Appendix A), confirming the robustness of the topological characterization beyond the reflection-matrix topological quantum number. 

The topological phase boundaries can be obtained analytically from the inverse localization length of the zero modes~\cite{Mondragon-ShemI2014}. For the generalized Bernoulli distribution defined above, the transition points satisfy
\begin{equation}
\label{Eq.3}
\prod_{j=1}^{M}\left|-t_1+\xi^{(j)}\right|^{p_j}=|t_2|.
\end{equation}
With the choice $t_2=1$, this condition reduces to $\prod_{j=1}^{M}\left|-t_1+\xi^{(j)}\right|^{p_j}=1$. Equivalently, the phase boundary can be written in logarithmic form as $\sum_{j=1}^{M}p_j\ln\left|-t_1+\xi^{(j)}\right|=\ln|t_2|$ (see Appendix A for details). We emphasize that this condition is not determined by a simple arithmetic average of the disordered intradimer hopping. Instead, it follows from the zero-mode recursion relation and reflects the multiplicative structure of the disordered couplings. This analytical condition accurately captures the boundaries of the disconnected topological windows obtained numerically.

\begin{figure*}[htbp]
\centering
\includegraphics[width=17cm]{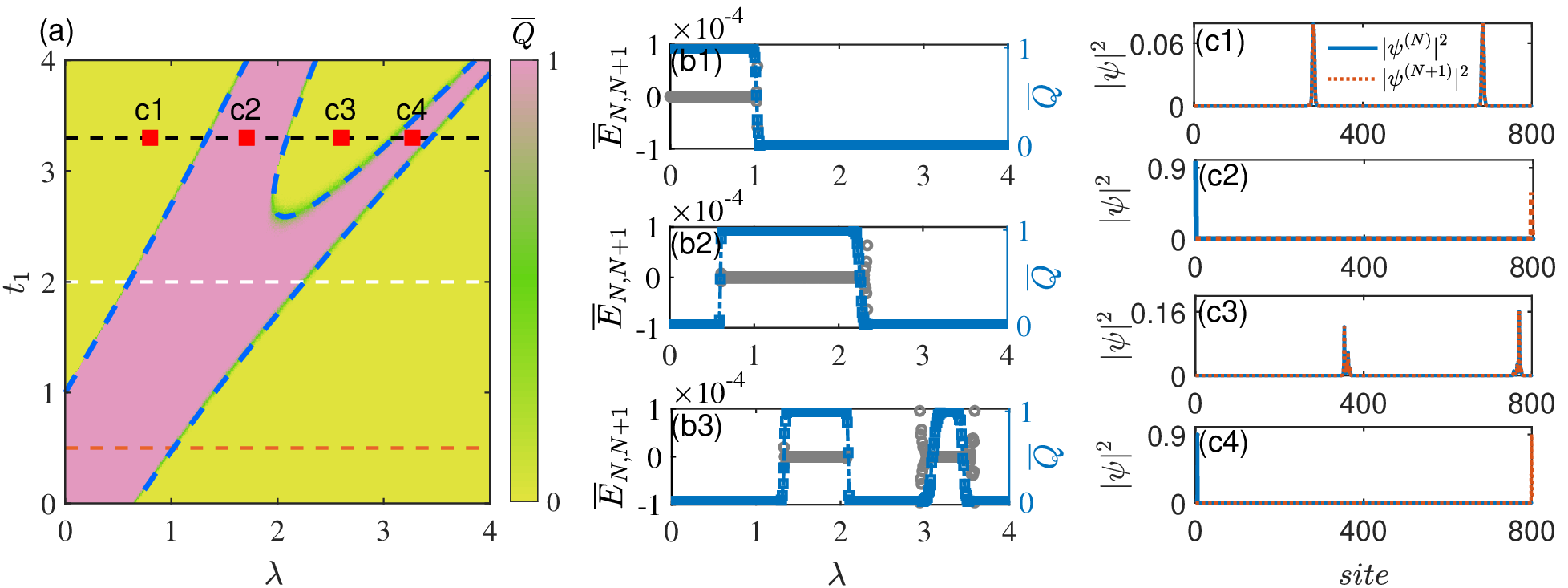}
\caption{(a) Disorder-averaged topological phase diagram characterized by $\overline{Q}$ as a function of the intradimer hopping $t_1$ and disorder amplitude $\lambda$ for $p_1=2/5$ and $p_2=3/5$. Orange, white, and black dashed lines indicate $t_1=0.5$, $2.0$, and $3.3$, respectively; blue dashed lines mark the analytical phase boundaries. (b1)--(b3) Disorder-averaged central energies $\overline{E}_N$, $\overline{E}_{N+1}$ and disorder-averaged topological quantum number $\overline{Q}$ versus $\lambda$ under OBCs for $t_1=0.5$, $2.0$, and $3.3$, respectively. (c1)--(c4) Density distributions of the $N$th and $(N+1)$th eigenstates under OBCs for $\lambda=0.80$, $1.71$, $2.60$, and $3.27$ [marked by red squares in (a)]. All data are averaged over $N_c=200$ disorder realizations, with $N=400$, $\xi^{(1)}=\lambda$, and $\xi^{(2)}=2\lambda$.}
\label{Fig1}
\end{figure*}

\begin{figure}[htbp]
\centering
\includegraphics[width=8.8cm]{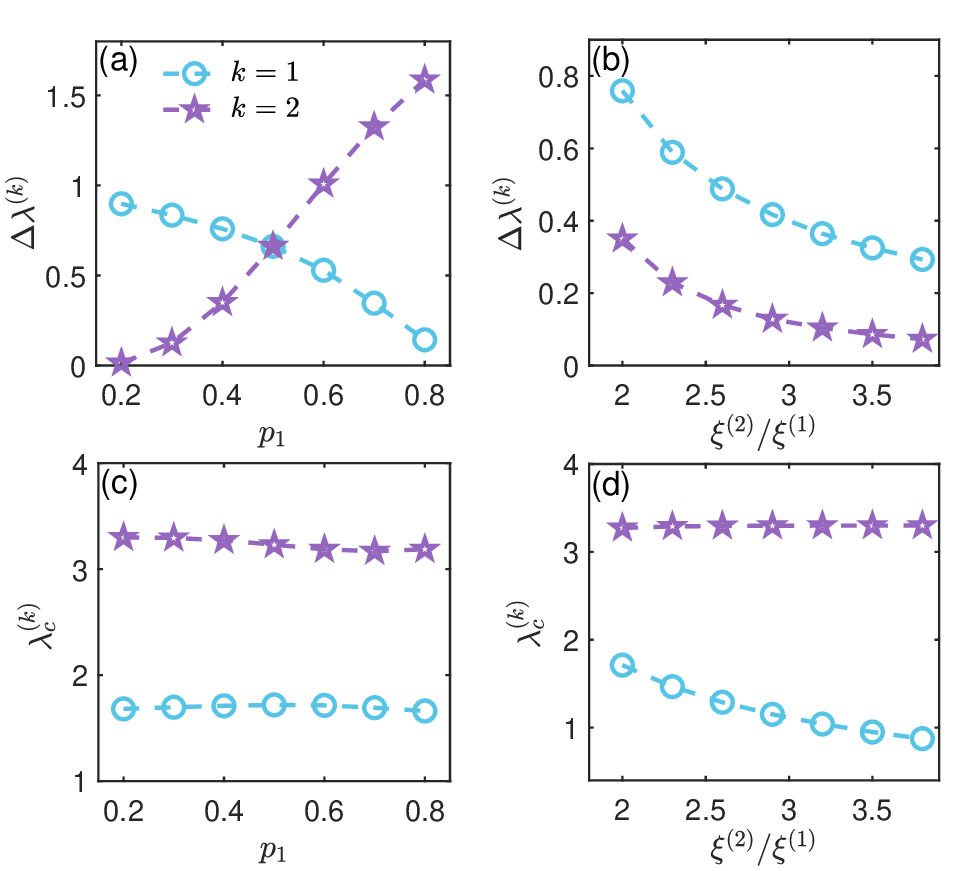}
\caption{Widths and center positions of the two disconnected topological windows for the binary generalized Bernoulli distribution. (a) Window widths $\Delta \lambda^{(k)}$ and (c) center positions $\lambda_c^{(k)}$ as functions of $p_1$ for $t_1=3.3$, $\xi^{(1)}=\lambda$, and $\xi^{(2)}=2\lambda$. (b) Window widths $\Delta \lambda^{(k)}$ and (d) center positions $\lambda_c^{(k)}$ as functions of the amplitude ratio $\xi^{(2)}/\xi^{(1)}$ for $t_1=3.3$, $\xi^{(1)}=\lambda$, $p_1=2/5$, and $p_2=3/5$.}
\label{Fig2}
\end{figure}

\begin{figure*}[htbp]
\centering
\includegraphics[width=17.3cm]{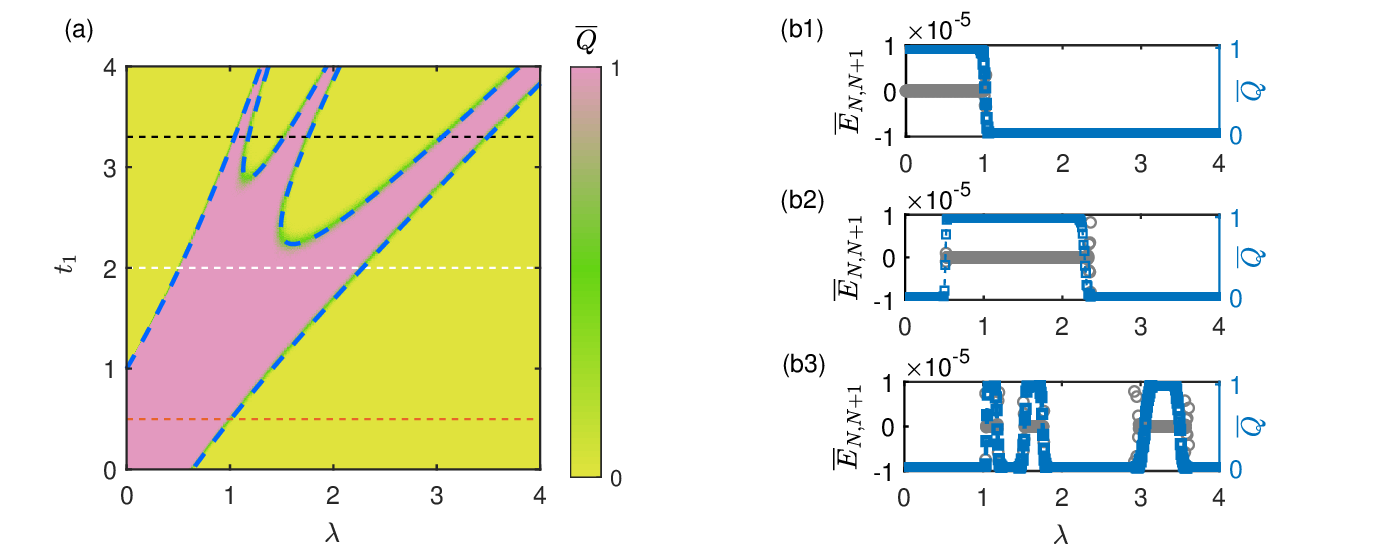}
\caption{(a) Disorder-averaged topological phase diagram characterized by $\overline{Q}$ as a function of intradimer hopping $t_1$ and disorder amplitude $\lambda$ for $p_1=1/2$ and $p_2=p_3=1/4$. Orange, white, and black dashed lines indicate $t_1=0.5$, $2.0$, and $3.3$; blue dashed lines mark the analytical phase boundaries. (b1)--(b3) Disorder-averaged central energies $\overline{E}_N$, $\overline{E}_{N+1}$, and disorder-averaged topological quantum number $\overline{Q}$ versus $\lambda$ under OBCs for $t_1=0.5$, $2.0$, and $3.3$, respectively. All data are averaged over $N_c=200$ disorder realizations, with $N=400$.}
\label{Fig3}
\end{figure*}

\section{Multiple reentrant topological windows}
%{\it Multiple reentrant topological windows.}
\begin{figure}[htbp]
	\centering
	\includegraphics[width=8.8cm]{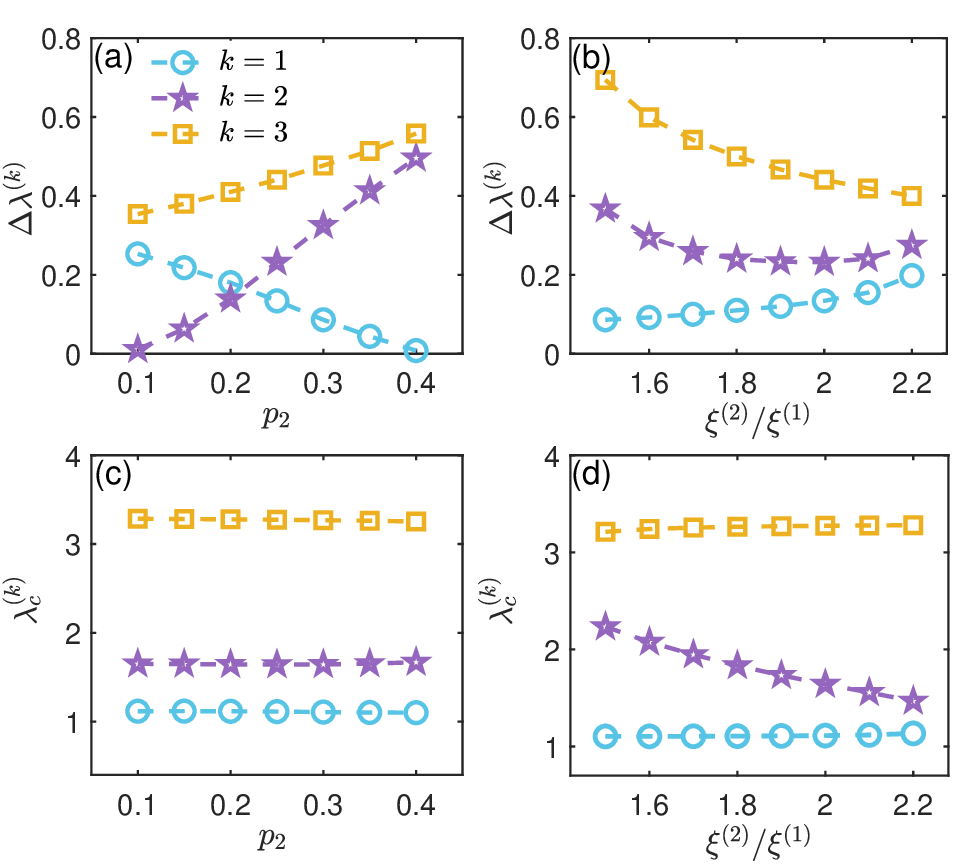}
	\caption{Widths and center positions of the three disconnected topological windows for the ternary generalized Bernoulli distribution. (a) Window widths $\Delta \lambda^{(k)}$ and (c) center positions $\lambda_c^{(k)}$ as functions of $p_2$ for $t_1=3.3$, $\xi^{(1)}=\lambda$, $\xi^{(2)}=2\lambda$, $\xi^{(3)}=3\lambda$, and $p_1=1/2$. (b) Window widths $\Delta \lambda^{(k)}$ and (d) center positions $\lambda_c^{(k)}$ as functions of the amplitude ratio $\xi^{(2)}/\xi^{(1)}$ for $t_1=3.3$, $\xi^{(1)}=\lambda$, $\xi^{(3)}=3\lambda$, $p_1=1/2$, and $p_2=p_3=1/4$.}
	\label{Fig4}
\end{figure}

We begin with the binary case, which corresponds to an SSH model with two-valued Bernoulli-type disorder. Figure~\ref{Fig1}(a) shows the disorder-averaged topological phase diagram as a function of $t_1$ and $\lambda$ for $\xi^{(1)}=\lambda$ and $\xi^{(2)}=2\lambda$ with probabilities $p_1=2/5$ and $p_2=3/5$, respectively. In the clean limit ($\lambda=0$), the system undergoes the usual topological transition at $t_1=1$, where $\overline{Q}$ changes from $1$ to $0$ as $t_1$ increases. When disorder is introduced into the nontrivial regime ($0<t_1<1$), the topological phase remains stable against weak disorder and becomes trivial only beyond a critical disorder strength. For $1<t_1<2.59$, increasing $\lambda$ drives the system from a trivial regime into a nontrivial window and then back into a trivial regime. For $t_1>2.59$, two disconnected nontrivial windows appear as $\lambda$ increases, indicating reentrant topological behavior. The analytical phase boundaries obtained from Eq.~\eqref{Eq.3} are shown by the blue dashed lines and agree well with the numerical phase diagram.

Figures~\ref{Fig1}(b1)-\ref{Fig1}(b3) display the disorder-averaged central energies $\overline{E}_N$ and $\overline{E}_{N+1}$ together with $\overline{Q}$ as functions of $\lambda$ under open boundary conditions (OBCs) for $t_1=0.5$, $2.0$, and $3.3$, respectively. Here, $\overline{E}_n=N_c^{-1}\sum_{c=1}^{N_c}E_n^{(c)}$, where $E_n^{(c)}$ is the $n$th eigenvalue for the $c$th disorder realization. For $t_1=0.5$ [Fig.~\ref{Fig1}(b1)], $\overline{Q}=1$ and both $\overline{E}_N$ and $\overline{E}_{N+1}$ remain pinned near zero in the weak-disorder regime, confirming the robustness of the nontrivial phase. For $t_1=2.0$ [Fig.~\ref{Fig1}(b2)], increasing $\lambda$ induces a nontrivial topological window in the interval $\lambda\in(0.60,2.25)$, where $\overline{Q}=1$ and a pair of nearly degenerate zero modes appears. For $t_1=3.3$ [Fig.~\ref{Fig1}(b3)], two disconnected topological windows occur at $\lambda\in(1.33,2.09)\cup(3.10,3.45)$, again accompanied by zero-energy boundary modes. These results show that the transitions of $\overline{Q}$ are consistently correlated with the appearance or disappearance of zero modes.

Figures~\ref{Fig1}(c1)-\ref{Fig1}(c4) show the density distributions of the $N$th and $(N+1)$th eigenstates, $|\psi^{(N)}|^2$ and $|\psi^{(N+1)}|^2$, for representative disorder realizations at $t_1=3.3$ and $\lambda=0.80$, $1.71$, $2.60$, and $3.27$. In the nontrivial windows, the two states are localized near the two ends of the chain, consistent with boundary zero modes. Outside these windows, the central states are no longer edge-localized and instead appear as bulk states for the disorder realizations shown here. Taken together, Fig.~\ref{Fig1} shows that binary generalized Bernoulli disorder can generate reentrant topological windows in the disordered SSH chain.

We next examine how the widths and positions of these disconnected topological windows depend on the parameters of the binary disorder distribution. For fixed $t_1$ and fixed distribution parameters, the phase boundaries are obtained from the real positive solutions of Eq.~\eqref{Eq.3}. After these transition points are ordered along the $\lambda$ axis, the topological windows are identified as the intervals between two neighboring transition points in which the topological quantum number takes the nontrivial value $\overline{Q}=1$. For the $k$th disconnected topological window, we denote its left and right boundaries by $\lambda_{L}^{(k)}$ and $\lambda_{R}^{(k)}$, respectively. The window width and center position are then defined as
\begin{equation}
\label{Eq.window}
\Delta\lambda^{(k)}=\lambda_{R}^{(k)}-\lambda_{L}^{(k)},\qquad
\lambda_c^{(k)}=\frac{\lambda_{L}^{(k)}+\lambda_{R}^{(k)}}{2}.
\end{equation}
Thus, the width $\Delta\lambda^{(k)}$ measures the extent of the $k$th topological interval in disorder strength, while the center $\lambda_c^{(k)}$ characterizes its location along the $\lambda$ axis. A change in $\lambda_c^{(k)}$ indicates a shift of the corresponding window, whereas a change in the separation between different centers, such as $\lambda_c^{(2)}-\lambda_c^{(1)}$, characterizes the relative motion of distinct windows.

Figure~\ref{Fig2}(a) shows the widths of the first and second topological windows for $t_1=3.3$ as functions of $p_1$, with $\xi^{(1)}=\lambda$ and $\xi^{(2)}=2\lambda$. Increasing $p_1$ causes the first window to shrink and the second to broaden. For $p_1<1/2$, the first window is wider than the second, whereas for $p_1>1/2$, the second becomes wider. This behavior follows directly from Eq.~\eqref{Eq.3}: changing $p_1$ redistributes the relative weights of the two disorder components in the logarithmic average of the hopping amplitudes. As a result, the transition points bounding each topological interval are shifted by different amounts. Since each width is the distance between two neighboring phase boundaries, these unequal shifts directly lead to the opposite width variations of the two windows. The corresponding center positions $\lambda_c^{(1)}$ and $\lambda_c^{(2)}$ are shown in Fig.~\ref{Fig2}(c). In contrast to the pronounced changes in the widths, both centers vary only weakly with $p_1$ over the parameter range considered. This indicates that changing the probabilities mainly redistributes the widths of the two disconnected topological windows, while their overall locations and relative separation along the $\lambda$ axis remain nearly unchanged.

Figure~\ref{Fig2}(b) shows the widths of the first and second topological windows as functions of the amplitude ratio $\xi^{(2)}/\xi^{(1)}$, with $\xi^{(1)}=\lambda$, $p_1=2/5$, and $p_2=3/5$. As $\xi^{(2)}/\xi^{(1)}$ increases, both window widths decrease, while the first window remains wider than the second throughout the parameter range shown. The corresponding center positions are shown in Fig.~\ref{Fig2}(d). In this case, increasing $\xi^{(2)}/\xi^{(1)}$ shifts the center of the first topological window toward smaller $\lambda$, whereas the center of the second window changes much more weakly. Therefore, the disorder probabilities mainly tune the relative widths of the windows, while the relative disorder amplitudes can also shift their positions in parameter space and modify their relative arrangement.

We next consider multivalued disorder distributions with $M>2$. Figure~\ref{Fig3}(a) shows the disorder-averaged topological phase diagram for $M=3$, with $\xi^{(1)}=\lambda$, $\xi^{(2)}=2\lambda$, and $\xi^{(3)}=3\lambda$, occurring with probabilities $p_1=1/2$ and $p_2=p_3=1/4$. For $t_1<1$, sufficiently strong disorder again destroys the nontrivial phase. For $1<t_1<2.23$, increasing $\lambda$ produces a single disconnected nontrivial window. For $t_1>2.23$, multiple disconnected topological windows appear as $\lambda$ increases. In the large-$t_1$ regime, the number of such windows matches the number of disorder components in the distribution, namely $M=3$. The analytical phase boundaries obtained from Eq.~\eqref{Eq.3} again agree with the numerical results. Figures~\ref{Fig3}(b1)-\ref{Fig3}(b3) show the disorder-averaged central energies $\overline{E}_N$ and $\overline{E}_{N+1}$ together with $\overline{Q}$ as functions of $\lambda$ for different values of $t_1$. As in the binary case, each transition of $\overline{Q}$ between $0$ and $1$ is accompanied by the appearance or disappearance of a pair of nearly degenerate zero modes, confirming the correspondence between the disorder-induced topological windows and boundary-state formation.

The widths and center positions of the topological windows for the ternary distribution are summarized in Fig.~\ref{Fig4}, using the same definitions as in Eq.~\eqref{Eq.window}. Figure~\ref{Fig4}(a) shows the widths of the first, second, and third topological windows for $t_1=3.3$ as functions of $p_2$, with $p_1=1/2$ fixed and $\xi^{(1)}=\lambda$, $\xi^{(2)}=2\lambda$, and $\xi^{(3)}=3\lambda$. As $\lambda$ increases, the three windows appear sequentially. Increasing $p_2$ causes the first window to narrow, while the second and third broaden. In particular, the first window is wider than the second for $p_2<0.215$, whereas the second becomes wider for $p_2>0.215$. This behavior again follows from the redistribution of weights in Eq.~\eqref{Eq.3}, which shifts different phase boundaries by different amounts. Since each window width is determined by the separation between neighboring transition points, unequal shifts of these boundaries lead directly to different width variations for the three windows. The corresponding center positions are shown in Fig.~\ref{Fig4}(c). Similar to the binary case, varying the disorder probabilities mainly changes the widths of the windows, while the center positions remain comparatively stable over the parameter range considered. This indicates that probability tuning primarily reshapes the disconnected topological intervals rather than strongly translating them as a whole.

Figure~\ref{Fig4}(b) further shows the widths of the three topological windows as functions of $\xi^{(2)}/\xi^{(1)}$, with $\xi^{(1)}=\lambda$, $\xi^{(3)}=3\lambda$, $p_1=1/2$, and $p_2=p_3=1/4$. As $\xi^{(2)}/\xi^{(1)}$ increases, the first window broadens, the third narrows, and the second displays a nonmonotonic dependence. Across the parameter range shown, the third window remains the widest, followed by the second and then the first. The corresponding center positions are shown in Fig.~\ref{Fig4}(d). In this case, increasing $\xi^{(2)}/\xi^{(1)}$ shifts the center of the second window toward smaller $\lambda$, whereas the other window centers change more weakly. These results further illustrate that the number, widths, and locations of the disconnected topological windows are jointly determined by the values and probabilities of the multivalued disorder distribution.

In Appendix A, we further present the disorder-averaged topological phase diagrams for $M=4$ and $M=5$. Their overall structures are similar to those for $M=2$ and $M=3$, while in the large-$t_1$ regime the number of disconnected topological windows increases with $M$. This trend supports the general picture that the complexity of the disorder distribution influences the number of reentrant topological windows in the disordered SSH model.

We also analyze the localization properties in Appendix B. Although reentrant extended regimes can appear in parts of the parameter space, their locations do not coincide one-to-one with the topological phase boundaries, indicating that the reentrant localization behavior and the reentrant topological transitions arise from different mechanisms.

\begin{figure}[htbp]
\centering
\includegraphics[width=8.8cm]{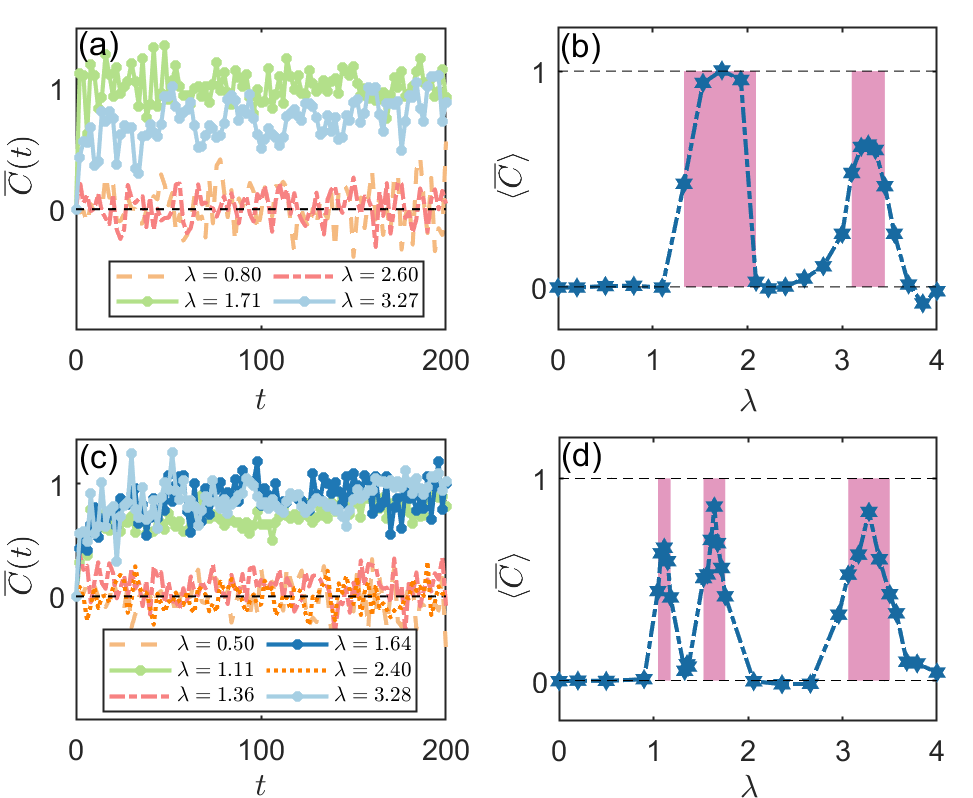}
\caption{Disorder-averaged mean chiral displacement $\overline{C}(t)$ versus time $t$ for various disorder amplitudes $\lambda$ for (a) $M=2$ and (c) $M=3$ ($N=200$ and $N_c=300$). Time-averaged mean chiral displacement $\langle \overline{C} \rangle$ as a function of $\lambda$ for (b) $M=2$ and (d) $M=3$ ($N=100$ and $N_c=300$), where the time average is taken from $t=0$ to $100$ with step size $0.5$. The pink regions indicate the topological windows identified from $\overline{Q}$. Parameters: [(a) and (b)] $\xi^{(1)}=\lambda$, $\xi^{(2)}=2\lambda$, $p_1=2/5$, $p_2=3/5$; [(c) and (d)] $\xi^{(1)}=\lambda$, $\xi^{(2)}=2\lambda$, $\xi^{(3)}=3\lambda$, $p_1=1/2$, $p_2=p_3=1/4$.}
\label{Fig5}
\end{figure}

\section{Dynamical detection}
%{\it Dynamical detection.} 
We finally discuss a dynamical probe of the disorder-induced topological transitions. The mean chiral displacement has been widely used to detect topological invariants in one-dimensional chiral-symmetric systems and remains useful in disordered settings after disorder averaging~\cite{MeierEJ2018,LuZ2025,LiX2024}. A possible photonic-waveguide implementation~\cite{ChristodoulidesD2003,LonghiS2009,GaranovichIL2012,CorrielliG2013,HuB2023} of the present model is outlined in Appendix C. In such a platform, the propagation distance plays the role of time, and the effective disordered intradimer hopping can be engineered through auxiliary waveguides.

To probe the topology dynamically, we consider the disorder-averaged mean chiral displacement, defined as
\begin{equation}\label{Eq.4}
\overline{C}(t)=\frac{2}{N_c}\sum_{c=1}^{N_c}\langle \Gamma X\rangle_c,
\end{equation}
where $\Gamma=I_N\otimes\sigma_z$ is the chiral-symmetry operator, with $\sigma_z$ the Pauli $z$ matrix and $I_N$ the $N\times N$ identity matrix, $X$ is the unit-cell position operator, and $\langle\cdots\rangle_c$ denotes the time-dependent expectation value for the $c$th disorder realization~\cite{LuZ2025,LiX2024,CardanoF2017}. 

Figures~\ref{Fig5}(a) and \ref{Fig5}(c) show the time evolution of the disorder-averaged mean chiral displacement for the binary and ternary disorder distributions, respectively. In both cases, $\overline{C}(t)$ remains close to zero in the trivial regimes, while it approaches values close to one inside the disconnected topological windows. For the binary case, $\lambda=0.80$ and $\lambda=2.60$ correspond to trivial regimes, whereas $\lambda=1.71$ and $\lambda=3.27$ lie inside the two topological windows. For the ternary case, $\lambda=0.50$, $1.36$, and $2.40$ correspond to trivial regimes, while $\lambda=1.11$, $1.64$, and $3.28$ lie inside the three disconnected topological windows.

To obtain a clearer indicator, we further consider the time-averaged mean chiral displacement $\langle\overline{C}\rangle$. In the ideal limit, the topological information is obtained from the long-time average of the mean chiral displacement. In practical numerical simulations and experiments, however, the accessible evolution or measurement time is always finite. Therefore, the oscillatory contributions from bulk states are not completely averaged out, and the resulting $\langle\overline{C}\rangle$ can deviate from the exact integer values expected in the long-time limit. This effect is more visible near topological phase boundaries, where the dynamical convergence is slower because the characteristic energy scale becomes small. By increasing the averaging time, especially for parameters well inside a trivial or nontrivial region, $\langle\overline{C}\rangle$ is expected to converge closer to the corresponding quantized value. Thus, the deviations of $\langle\overline{C}\rangle$ from exact integers mainly reflect the finite-time nature of the dynamical measurement rather than a change in the topological classification.

Figures~\ref{Fig5}(b) and \ref{Fig5}(d) show that, as $\lambda$ increases, $\langle\overline{C}\rangle$ changes between values near zero and near one in parameter intervals consistent with the topological windows identified from $\overline{Q}$. In particular, two such nontrivial windows are resolved in Fig.~\ref{Fig5}(b) for $M=2$, and three are resolved in Fig.~\ref{Fig5}(d) for $M=3$. The locations of these dynamical crossovers are consistent with the phase boundaries obtained from $\overline{Q}$ and Eq.~\eqref{Eq.3}. These results show that the time-averaged mean chiral displacement provides a useful dynamical signature of the disorder-induced topological transitions and of the disconnected reentrant topological windows in our model.

\section{Conclusions}

We have investigated reentrant topological behavior in a one-dimensional SSH model with generalized Bernoulli-type disorder in the intradimer hopping amplitudes. We showed that a multivalued discrete disorder distribution can split the nontrivial regime into multiple disconnected topological windows as the disorder strength is varied. The phase boundaries are obtained analytically from the inverse localization length of the zero modes and are governed by the weighted geometric mean of the absolute values of the disordered hopping amplitudes. The analytical predictions agree well with numerical results based on the reflection-matrix topological quantum number and the real-space winding number.

We further showed that the number, widths, and locations of the disconnected topological windows can be tuned by the number of disorder components, their probabilities, and their relative amplitudes. In particular, increasing the number of distinct disorder values can increase the number of reentrant topological windows in appropriate parameter regimes, while changing the probabilities and relative amplitudes modifies the widths and center positions of the windows. We also found that this reentrant topological behavior is not restricted to intradimer disorder: the complementary case with generalized Bernoulli disorder in the interdimer hopping exhibits the same qualitative phenomenology, with complementary phase diagrams.

In addition, we demonstrated that the mean chiral displacement provides a useful dynamical probe of the disorder-induced topological transitions. The deviations of the time-averaged mean chiral displacement from exact integer values can be attributed mainly to finite-time averaging in numerical simulations and realistic measurements. Rather than introducing new topological classes, generalized Bernoulli disorder reshapes the nontrivial regime into multiple disconnected windows in parameter space. Our results clarify how the statistical structure of a multivalued disorder distribution governs the emergence and tunability of reentrant topological windows in one-dimensional chiral lattices.

\section*{ACKNOWLEDGMENTS}

Z. X. is supported by Quantum Science and Technology-National Science and Technology Major Project (Grant No. 2025ZD0300400), the NSFC (Grant No. 12375016), and Beijing National Laboratory for Condensed Matter Physics (No. 2023BNLCMPKF001). Y. Z. is supported by the NSFC (Grant No. 12474492 and No. 12461160324) and the Challenge Project of Nuclear Science (Grant No. TZ2025017). S. C. is supported by National Key Research and Development Program of China (Grant No. 2023YFA1406704), the NSFC under Grants No. 12174436 and No. T2121001, and the Strategic Priority Research Program of Chinese Academy of Sciences under Grant No. XDB33000000.

\section*{DATA AVAILABILITY}
%{\it Data availability.} 
The data that support the findings of this article are not publicly available. The data are available from the authors upon reasonable request.

\appendix
\section{TOPOLOGICAL CHARACTERIZATION}

\subsection{Reflection-matrix topological quantum number}
We consider a dimerized polymer chain such as polyacetylene~\cite{FulgaIC2011}, with alternating long and short bonds, described by the SSH Hamiltonian
\begin{equation}
    \begin{aligned}
	H &=-\sum_{n=1}^{2N}\left(\psi_{n+1}^{\dagger}t_{n+1,n}\psi_{n}+\textrm{H.c.}\right) \\
	t_{n+1,n} &=
	\begin{cases}
		t_n^{(1)}, & \text{intradimer hopping}, \\
		t_n^{(2)}, & \text{interdimer hopping}.
	\end{cases}
    \end{aligned}
	\label{eq1}
\end{equation}
Here, $t_{n+1,n}$ denotes the nearest-neighbor hopping amplitude, $\psi_n$ is the wave amplitude at site $n$, and $2N$ is the total number of lattice sites in the chain. To characterize the topology of the SSH chain, we consider the scattering matrix $S$ at zero energy, which relates the incoming and outgoing wave amplitudes~\cite{Zhang2016},
\begin{equation}
	S=
	\begin{pmatrix}
		\tilde{R}_{\leftarrow} & \tilde{T}_{\leftarrow} \\
		\tilde{T}_{\rightarrow} & \tilde{R}_{\rightarrow}
	\end{pmatrix},
	\label{eq2}
\end{equation}
where $\tilde{R}_{\leftarrow}$ and $\tilde{R}_{\rightarrow}$ are the reflection amplitudes from the left and right ends of the chain, respectively, and $\tilde{T}_{\leftarrow}$ and $\tilde{T}_{\rightarrow}$ are the corresponding transmission amplitudes. The $\mathcal{Z}_2$ topological quantum number is defined as
\begin{equation}
	\mathcal{Q}=\mathrm{sgn}\!\left(\tilde{R}_{\leftarrow}\right)
	=\mathrm{sgn}\!\left(\tilde{R}_{\rightarrow}\right),
	\label{eq3}
\end{equation}
where $\mathrm{sgn}(\cdots)$ denotes the sign function. The nontrivial regime with zero-energy end states corresponds to $\mathcal{Q}=-1$.

The scattering matrix can be obtained from the transfer-matrix approach. Based on the Hamiltonian in Eq.~(\ref{eq1}), the zero-energy Schr\"odinger equation gives
\begin{equation}
	\begin{pmatrix}
		t_{n+1,n}\psi_n \\
		\psi_{n+1}
	\end{pmatrix}
	=
	\tilde{\mathcal{M}}_n
	\begin{pmatrix}
		t_{n,n-1}\psi_{n-1} \\
		\psi_n
	\end{pmatrix},
	\label{eq4}
\end{equation}
with
\begin{equation}
	\tilde{\mathcal{M}}_n=
	\begin{pmatrix}
		0 & t_{n+1,n} \\
		-1/t_{n+1,n} & 0
	\end{pmatrix}.
	\label{eq5}
\end{equation}
The wave amplitudes at the two ends of the chain are connected by the total transfer matrix
\begin{equation}
	\tilde{\mathcal{M}}
	=
	\tilde{\mathcal{M}}_{2N}\tilde{\mathcal{M}}_{2N-1}\cdots \tilde{\mathcal{M}}_2\tilde{\mathcal{M}}_1
	=
	\begin{pmatrix}
		X & 0 \\
		0 & 1/X
	\end{pmatrix},
	\label{eq6}
\end{equation}
where
\begin{equation}
	X=(-1)^N\prod_{n=1}^{N}\frac{t_n^{(2)}}{t_n^{(1)}}.
	\label{eq7}
\end{equation}

To obtain the scattering matrix, we transform from the site basis to the basis of right- and left-moving waves. The transfer matrix transforms as
\begin{equation}
	\mathcal{M}_n=U^{T}\tilde{\mathcal{M}}_nU^{*},
	\label{eq8}
\end{equation}
where
\begin{equation}
	U=\frac{1}{\sqrt{2}}
	\begin{pmatrix}
		1 & 1 \\
		i & -i
	\end{pmatrix}.
	\label{eq9}
\end{equation}
The total transfer matrix in this basis is
\begin{equation}
	\mathcal{M}
	=
	\mathcal{M}_{2N}\mathcal{M}_{2N-1}\cdots \mathcal{M}_2\mathcal{M}_1
	=
	\frac{1}{2X}
	\begin{pmatrix}
		X^2+1 & X^2-1 \\
		X^2-1 & X^2+1
	\end{pmatrix}.
	\label{eq10}
\end{equation}
The reflection amplitudes $(\tilde{R}_{\leftarrow},\tilde{R}_{\rightarrow})$ and transmission amplitudes $(\tilde{T}_{\leftarrow},\tilde{T}_{\rightarrow})$ are then determined from
\begin{equation}
	\begin{pmatrix}
		\tilde{T}_{\rightarrow} \\
		0
	\end{pmatrix}
	=
	\mathcal{M}
	\begin{pmatrix}
		1 \\
		\tilde{R}_{\leftarrow}
	\end{pmatrix},
	\qquad
	\begin{pmatrix}
		\tilde{R}_{\rightarrow} \\
		1
	\end{pmatrix}
	=
	\mathcal{M}
	\begin{pmatrix}
		0 \\
		\tilde{T}_{\leftarrow}
	\end{pmatrix}.
	\label{eq11}
\end{equation}
Solving these relations gives
\begin{equation}
	\tilde{R}_{\leftarrow}
	=
	\frac{1-X^2}{1+X^2},
	\label{eq12}
\end{equation}
and therefore the $\mathcal{Z}_2$ topological quantum number can be written as
\begin{equation}
	\begin{split}
	\mathcal{Q}
	&=
	\mathrm{sgn}\!\left(\tilde{R}_{\leftarrow}\right)
	=
	\mathrm{sgn}\!\left(\frac{1-X^2}{1+X^2}\right) \\
	&=
	\mathrm{sgn}\!\left[\prod_{n=1}^{N}\left(t_n^{(1)}\right)^2-\prod_{n=1}^{N}\left(t_n^{(2)}\right)^2\right].
    \end{split}
	\label{eq13}
\end{equation}

To connect with the notation used in the main text, we further define
\begin{equation}
	Q=\frac{1}{2}(1-\mathcal{Q}),
	\label{eq14}
\end{equation}
so that $Q=1$ corresponds to the nontrivial regime and $Q=0$ to the trivial regime. For the model studied in the main text, this becomes
\begin{equation}
	Q=\frac{1}{2}\left(1-\mathrm{sgn}\left[\prod_i\left(-t_1+\xi_i\right)^2-\left[ t_2 \right]^{2N} \right]\right).
	\label{eq15}
\end{equation}

\subsection{Real-space winding number}

\begin{figure*}[htbp]
	\centering	
	\includegraphics[width=13.5cm]{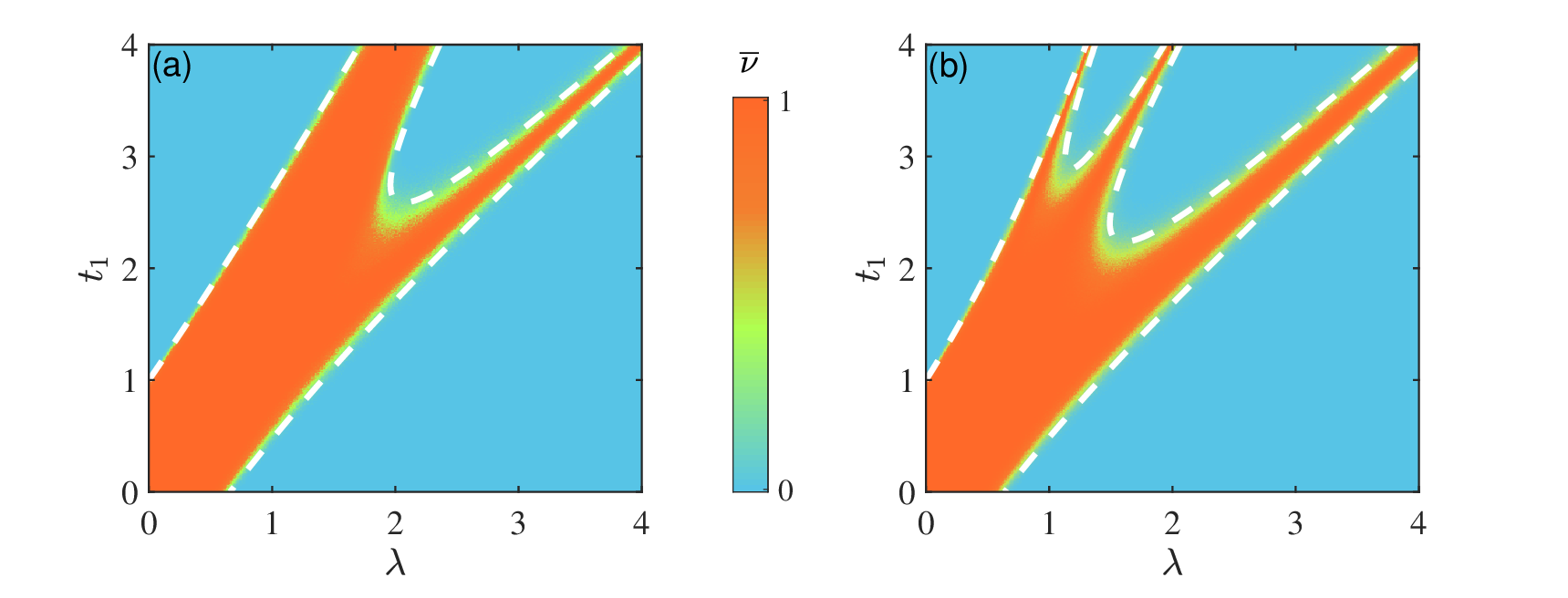}
	\caption{Phase diagrams characterized by the disorder-averaged real-space winding number $\overline{\nu}$ as functions of the intradimer hopping amplitude $t_1$ and disorder amplitude $\lambda$ for (a) $p_1=2/5$, $p_2=3/5$, $\xi^{(1)}=\lambda$, and $\xi^{(2)}=2\lambda$; (b) $p_1=1/2$, $p_2=p_3=1/4$, $\xi^{(1)}=\lambda$, $\xi^{(2)}=2\lambda$, and $\xi^{(3)}=3\lambda$. The white dashed lines indicate the analytical phase boundaries. All data are averaged over $N_c=50$ disorder realizations.}
	\label{FigS1}
\end{figure*}

\begin{figure*}[htbp]
	\centering
	\includegraphics[width=12.5cm]{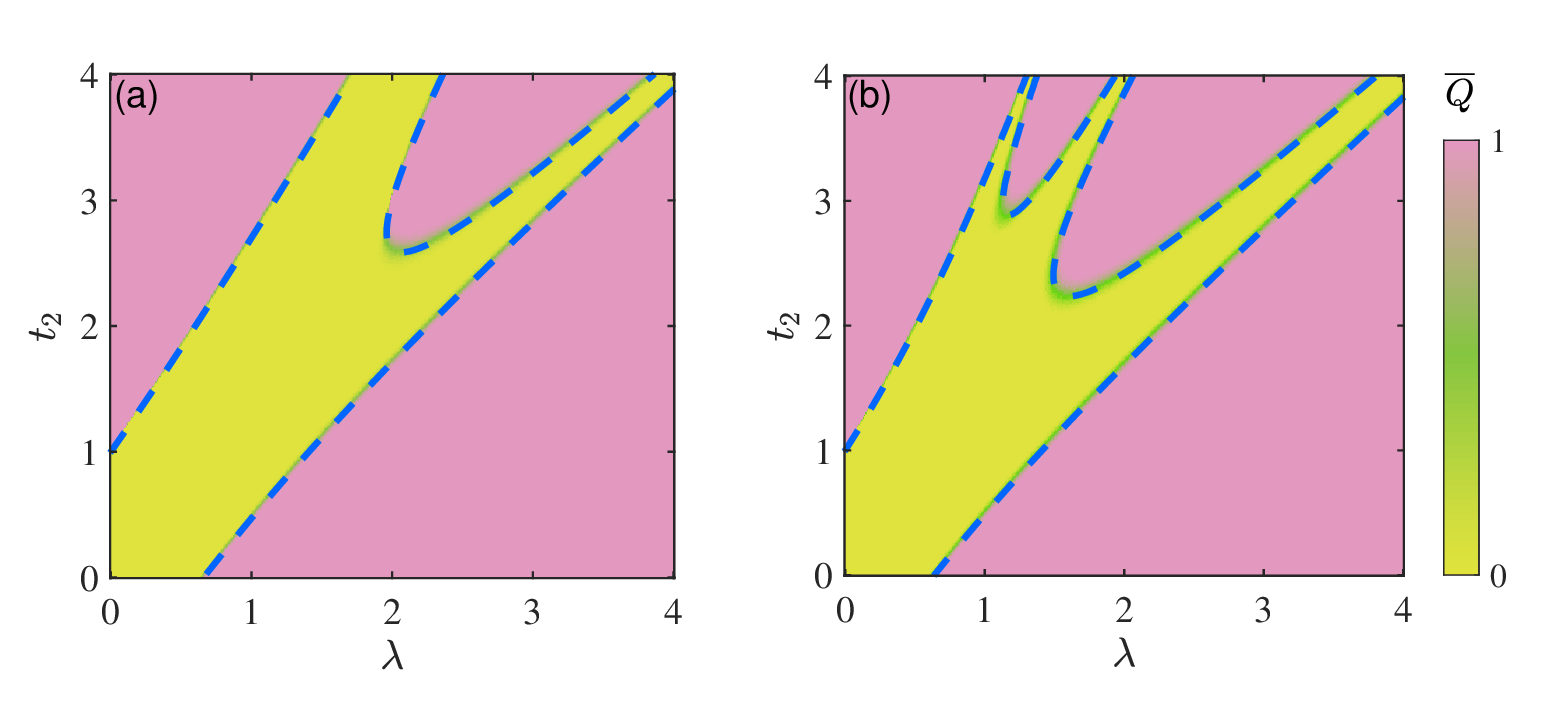}
	\caption{Disorder-averaged topological phase diagrams as functions of the interdimer hopping amplitude $t_2$ and disorder strength $\lambda$ for (a) $\xi^{(1)}=\lambda$, $\xi^{(2)}=2\lambda$, $p_1=2/5$, and $p_2=3/5$; (b) $\xi^{(1)}=\lambda$, $\xi^{(2)}=2\lambda$, $\xi^{(3)}=3\lambda$, $p_1=1/2$, and $p_2=p_3=1/4$. The blue dashed lines indicate the analytical phase boundaries given by Eq.~(\ref{Eq.31}). All data are averaged over $N_c=200$ disorder realizations.}
	\label{FigS2}
\end{figure*}
\begin{figure*}[htbp]
	\centering
	\includegraphics[width=13cm]{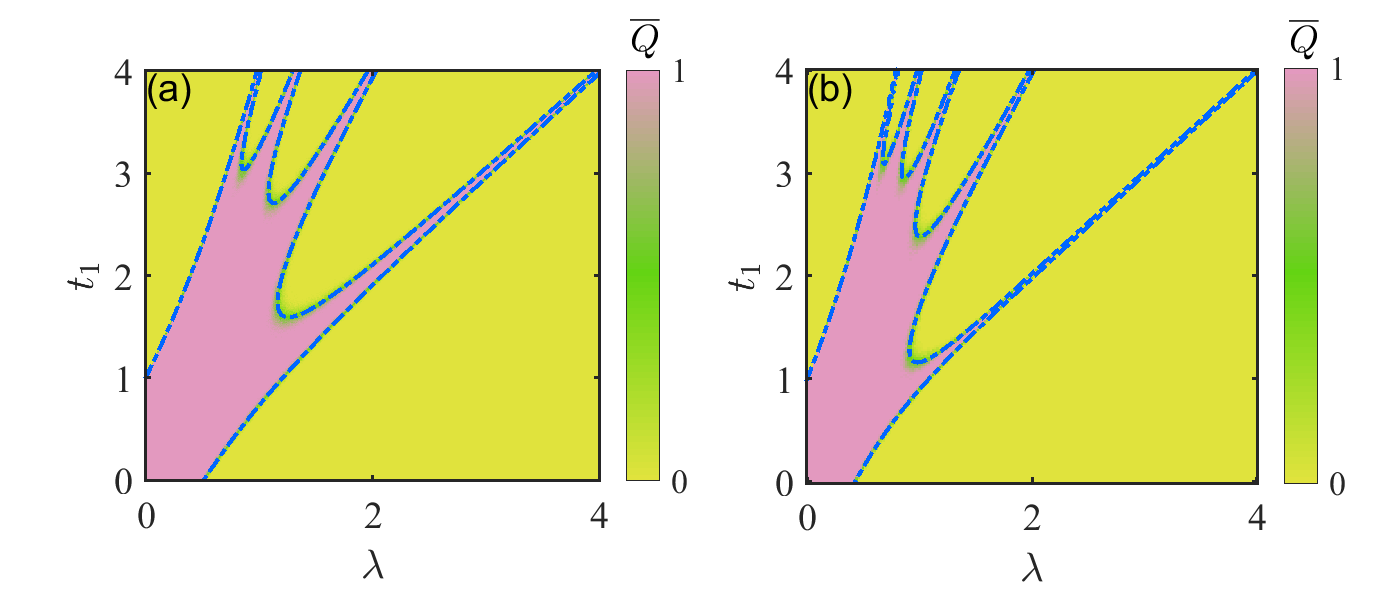}
	\caption{Disorder-averaged topological phase diagrams as functions of the intradimer hopping amplitude $t_1$ and disorder strength $\lambda$ for (a) $\xi^{(m)}=m\lambda$ with $m=1,2,3,4$, $p_1=7/20$, $p_2=1/4$, and $p_3=p_4=1/5$; (b) $\xi^{(m)}=m\lambda$ with $m=1,2,3,4,5$, $p_1=p_2=1/4$, and $p_3=p_4=p_5=1/6$. The blue dashed lines indicate the analytical phase boundaries. All data are averaged over $N_c=200$ disorder realizations.}
	\label{FigS3}
\end{figure*}

As a complementary topological characterization, we also compute the phase diagram using the real-space winding number. This quantity remains well defined in the absence of translational symmetry for a system with chiral symmetry~\cite{RoyK2024,RoyS2023,WangXM2025} and is given by
\begin{equation}
	\label{Eq.16}
	\nu=\frac{1}{L'}\mathrm{Tr}'\left(\Gamma \tilde{Q}\left[\tilde{Q},\tilde{X}\right]\right),
\end{equation}
where $\Gamma$ is the chiral-symmetry operator, $\tilde{X}$ is the position operator, and $\mathrm{Tr}'$ denotes the trace restricted to a central region of length $L'$. Following the standard numerical implementation, $\tilde{Q}$ is constructed from the positive- and negative-energy eigenstates of the Hamiltonian,
\begin{equation}
	\tilde{Q}=\sum_{E_{\tilde{j}}>0}|\tilde{j}\rangle\langle \tilde{j}|-\sum_{E_{\tilde{j}}<0}|\tilde{j}\rangle\langle \tilde{j}|,
\end{equation}
which is the real-space representation of the flattened Hamiltonian. To reduce sample-to-sample fluctuations, we further define the disorder-averaged real-space winding number as $\overline{\nu}=N_c^{-1}\sum_{c=1}^{N_c}\nu_c$, where $N_c$ is the total number of disorder realizations and $\nu_c$ is the real-space winding number for the $c$th realization. Figure~\ref{FigS1} shows the phase diagrams characterized by $\overline{\nu}$. The phase boundaries obtained from $\overline{\nu}$ agree with those obtained from the disorder-averaged topological quantum number $\overline{Q}$ in the main text within numerical resolution.

\subsection{Analytical phase boundaries}

The topological phase boundaries can be obtained from the inverse localization length of the zero modes. In the topologically nontrivial regime, the zero-energy modes are exponentially localized near the ends of the chain. At the transition to the trivial regime, the corresponding localization length diverges~\cite{Mondragon-ShemI2014}. Therefore, the phase boundaries are determined by the condition that the inverse localization length vanishes.

For the modulated SSH model in the main text, the zero-energy Schr\"odinger equation $\hat{H}|\psi\rangle=0$ gives the recursion relations
\begin{equation}
	\label{Eq.18}
	\begin{split}
		\left(-t_1+\xi_i\right)\psi_{i,B}-t_2\psi_{i-1,B}=0, \\
		\left(-t_1+\xi_i\right)\psi_{i,A}-t_2\psi_{i+1,A}=0,
	\end{split}
\end{equation}
where $\psi_{i,\sigma}$ is the zero-mode amplitude on sublattice $\sigma=A,B$ in the $i$th unit cell. Iterating the equation for the $A$ sublattice yields
\begin{equation}
	\psi_{N+1,A}=(-1)^N\prod_{i=1}^{N}\frac{-t_1+\xi_i}{-t_2}\,\psi_{1,A}.
	\label{Eq.19}
\end{equation}
The inverse localization length of the zero mode is then defined as
\begin{equation}
	\gamma=-\lim_{N\to\infty}\frac{1}{N}\ln\left|\frac{\psi_{N+1,A}}{\psi_{1,A}}\right|.
	\label{Eq.20}
\end{equation}
Substituting Eq.~(\ref{Eq.19}) into Eq.~(\ref{Eq.20}) gives
\begin{equation}
	\label{Eq.21}
	\begin{split}
		\gamma
		&=-\lim_{N\to\infty}\frac{1}{N}\ln\left|\prod_{i=1}^{N}\frac{-t_1+\xi_i}{-t_2}\right|.
	\end{split}
\end{equation}
For a generalized Bernoulli distribution with values $\xi^{(1)},\xi^{(2)},\ldots,\xi^{(M)}$ occurring with probabilities $p_1,p_2,\ldots,p_M$, let $l_j$ denote the number of occurrences of $\xi^{(j)}$ in $N$ unit cells, so that $\sum_{j=1}^{M}l_j=N$. In the thermodynamic limit, $l_j/N\to p_j$, and Eq.~(\ref{Eq.21}) becomes
\begin{equation}
	\label{Eq.22}
	\gamma
	=\ln|t_2|-\sum_{j=1}^{M}p_j\ln\left|-t_1+\xi^{(j)}\right|.
\end{equation}
The topological phase boundaries are therefore determined by the condition $\gamma=0$, which yields
\begin{equation}
	\label{Eq.23}
	\sum_{j=1}^{M}p_j\ln\left|-t_1+\xi^{(j)}\right|=\ln|t_2|.
\end{equation}
Equivalently, this can be written as
\begin{equation}
	\label{Eq.24}
	\prod_{j=1}^{M}\left|-t_1+\xi^{(j)}\right|^{p_j}=|t_2|.
\end{equation}
With the choice $t_2=1$ used in the main text, Eq.~(\ref{Eq.24}) reduces to $\prod_{j=1}^{M}\left|-t_1+\xi^{(j)}\right|^{p_j}=1$. This result shows that the phase boundaries are controlled by the weighted geometric mean of the absolute values of the disordered intradimer hopping amplitudes, rather than by a simple arithmetic average.

\subsection{Generalized Bernoulli disorder in the interdimer hopping}

In this section, we consider the complementary case in which the generalized Bernoulli disorder is introduced in the interdimer hopping, while the intradimer hopping remains uniform. The corresponding Hamiltonian is
\begin{equation}
	\label{Eq.25}
	\hat{H}'=-\sum_i\left[t_1\hat{c}_{i,A}^{\dagger}\hat{c}_{i,B}+(t_2-\xi_i)\hat{c}_{i,B}^{\dagger}\hat{c}_{i+1,A}+\textrm{H.c.}\right],
\end{equation}
where $\hat{c}_{i,\sigma}$ ($\hat{c}_{i,\sigma}^{\dagger}$) annihilates (creates) a particle on sublattice $\sigma=A,B$ of the $i$th unit cell. The random variable $\xi_i$ is independently drawn from a generalized Bernoulli distribution with $M$ possible values $\xi^{(1)},\xi^{(2)},\ldots,\xi^{(M)}$ and corresponding probabilities $p_1,p_2,\ldots,p_M$, satisfying $\sum_{j=1}^{M}p_j=1$. Throughout this section, we set $t_1=1$ as the energy unit.

As in the intradimer-disorder case discussed in the main text, the topological phase boundaries can be obtained from the inverse localization length of the zero modes. For the zero-energy state satisfying $\hat{H}'|\psi\rangle=0$, the Schr\"odinger equation gives
\begin{equation}
	\label{Eq.26}
	\begin{split}
		-t_1\psi_{i,B}-(t_2-\xi_{i-1})\psi_{i-1,B}=0, \\
		-t_1\psi_{i,A}-(t_2-\xi_i)\psi_{i+1,A}=0,
	\end{split}
\end{equation}
where $\psi_{i,\sigma}$ denotes the zero-mode amplitude on sublattice $\sigma=A,B$ in the $i$th unit cell. Iterating the equation for the $A$ sublattice yields
\begin{equation}
	\label{Eq.27}
	\psi_{N+1,A}=(-1)^N\prod_{i=1}^{N}\frac{t_1}{t_2-\xi_i}\,\psi_{1,A}.
\end{equation}
The inverse localization length is then defined as
\begin{equation}
	\label{Eq.28}
	\gamma=-\lim_{N\to\infty}\frac{1}{N}\ln\left|\frac{\psi_{N+1,A}}{\psi_{1,A}}\right|.
\end{equation}
Substituting Eq.~(\ref{Eq.27}) into Eq.~(\ref{Eq.28}), we obtain
\begin{equation}
	\label{Eq.29}
	\begin{split}
		\gamma
		&=-\lim_{N\to\infty}\frac{1}{N}\ln\left|\prod_{i=1}^{N}\frac{t_1}{t_2-\xi_i}\right| \\
		&=\sum_{j=1}^{M}p_j\ln\left|t_2-\xi^{(j)}\right|-\ln|t_1|.
	\end{split}
\end{equation}
The topological phase boundaries are determined by the condition $\gamma=0$, which gives
\begin{equation}
	\label{Eq.30}
	\sum_{j=1}^{M}p_j\ln\left|t_2-\xi^{(j)}\right|=\ln|t_1|.
\end{equation}
Equivalently,
\begin{equation}
	\label{Eq.31}
	\prod_{j=1}^{M}\left|t_2-\xi^{(j)}\right|^{p_j}=|t_1|.
\end{equation}
With the choice $t_1=1$, this reduces to $\prod_{j=1}^{M}\left|t_2-\xi^{(j)}\right|^{p_j}=1$. Equation~(\ref{Eq.31}) shows that the phase boundary is controlled by the weighted geometric mean of the disordered interdimer hopping amplitudes, rather than by their arithmetic average.

The resulting disorder-averaged topological phase diagrams for binary and ternary generalized Bernoulli distributions are shown in Figs.~\ref{FigS2}(a) and \ref{FigS2}(b), respectively. Compared with the corresponding phase diagrams for generalized Bernoulli disorder in the intradimer hopping under the same parameters in the main text, these phase diagrams exhibit a complementary structure: the topologically trivial regions in Fig.~\ref{FigS2}(a) [Fig.~\ref{FigS2}(b)] coincide with the topologically nontrivial regions in Fig.~\ref{Fig1}(a) [Fig.~\ref{Fig3}(a)], and vice versa. This behavior follows directly from the phase-boundary condition, since introducing the disorder into the interdimer hopping effectively interchanges the roles of the uniform and disordered couplings in the zero-mode recursion relation. Therefore, although the topological character of the corresponding parameter regions is reversed, the qualitative phenomenology remains the same. In particular, a multivalued Bernoulli distribution in the interdimer hopping can also generate disconnected reentrant topological windows as the disorder strength is varied.

\subsection{Additional examples for $M=4$ and $M=5$}
To further illustrate how the number of disorder components $M$ affects the reentrant topological structure, we present in Fig.~\ref{FigS3} the disorder-averaged topological phase diagrams as functions of $t_1$ and $\lambda$ for the cases $M=4$ and $M=5$. For $M=4$, we take $\xi^{(m)}=m\lambda$ with $m=1,2,3,4$, and the corresponding probabilities are chosen as $p_1=7/20$, $p_2=1/4$, $p_3=1/5$, and $p_4=1/5$. For $M=5$, we use $\xi^{(m)}=m\lambda$ with $m=1,2,3,4,5$, with probabilities $p_1=1/4$, $p_2=1/4$, $p_3=1/6$, $p_4=1/6$, and $p_5=1/6$.

As shown in Fig.~\ref{FigS3}, the overall structure of the phase diagrams remains consistent with the cases $M=2$ and $M=3$ discussed in the main text. In particular, in the large-$t_1$ regime, the number of disconnected topological windows increases with $M$. For the representative parameter sets considered here, four distinct topological windows are observed for $M=4$, while five such windows appear for $M=5$. These results further support the general trend that increasing the number of values in the generalized Bernoulli distribution enriches the reentrant topological structure and increases the number of disconnected topological intervals.

Therefore, higher-valued generalized Bernoulli disorder provides a simple and flexible way to engineer multiple reentrant topological windows in one-dimensional chiral lattices. The examples for $M=4$ and $M=5$ shown here extend the results in the main text and further demonstrate the tunability of the topological phase diagram through the choice of disorder values and their associated probabilities.

\section{LOCALIZATION PROPERTIES}

\begin{figure*}[htbp]
	\centering
	\includegraphics[width=18cm]{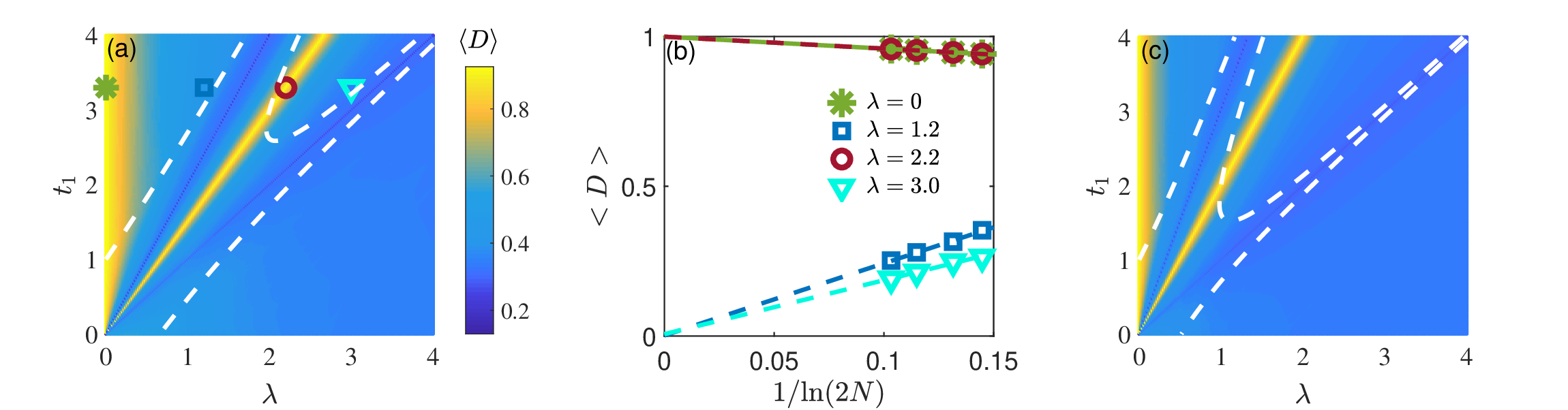}
	\caption{Mean fractal dimension $\langle D \rangle$ as a function of the intradimer hopping amplitude $t_1$ and disorder strength $\lambda$ for (a) $\xi^{(1)}=\lambda$, $\xi^{(2)}=2\lambda$, $p_1=2/5$, and $p_2=3/5$; (c) $\xi^{(1)}=\lambda$, $\xi^{(2)}=3\lambda$, $p_1=2/5$, and $p_2=3/5$. (b) Finite-size scaling of the mean fractal dimension $\langle D\rangle$ for different values of $\lambda$. The markers in (b) correspond to the same parameter sets indicated by the respective symbols in (a). The white dashed lines in (a) and (c) denote the topological phase boundaries.}
	\label{FigS4}
\end{figure*}

\begin{figure*}[htbp]
	\centering
	\includegraphics[width=13cm]{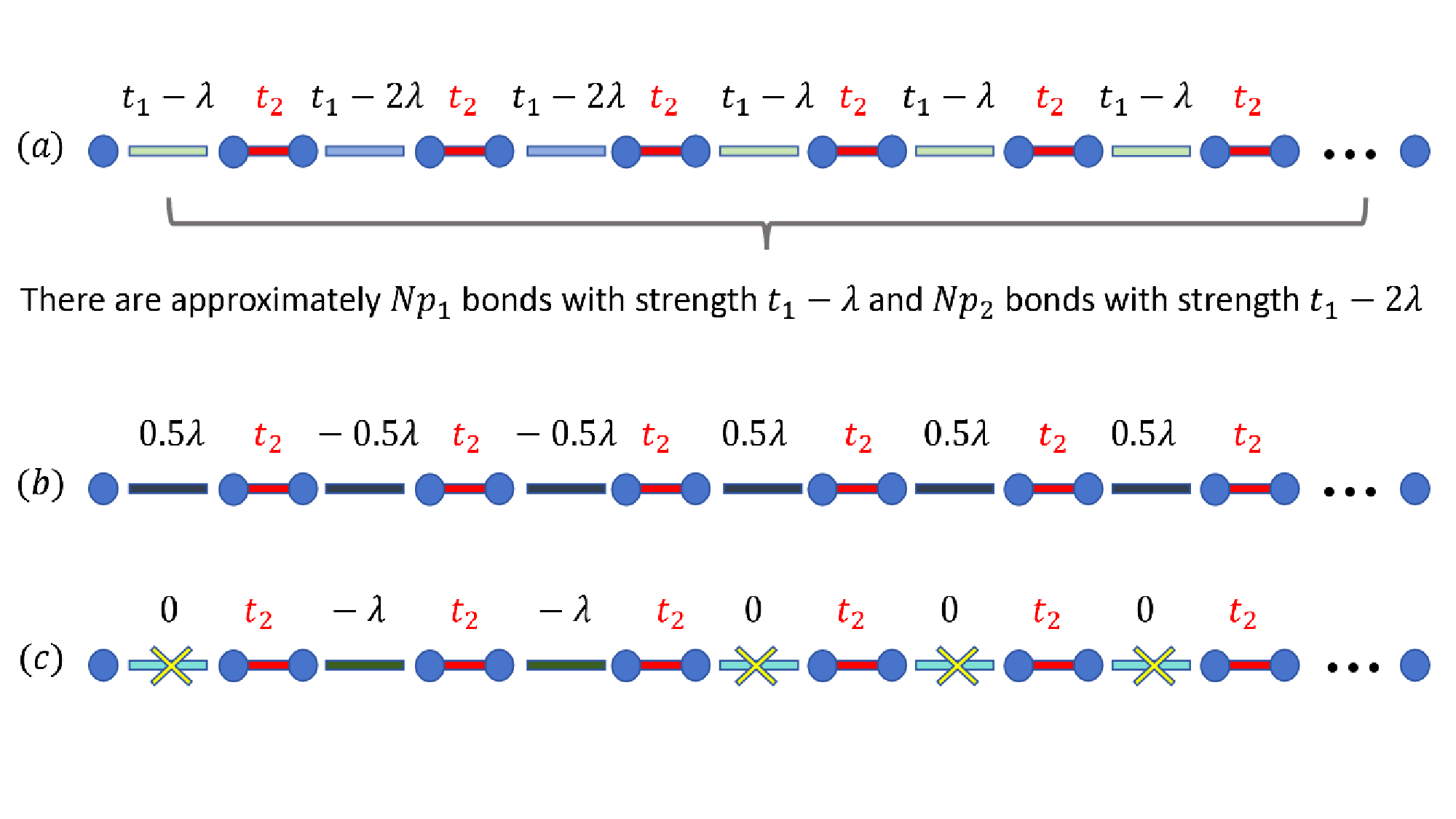}
	\caption{Schematic illustration of the SSH chain with binary generalized Bernoulli disorder under different parameter conditions.}
	\label{FigS5}
\end{figure*}

\begin{figure*}[htbp]
	\centering
	\includegraphics[width=17cm]{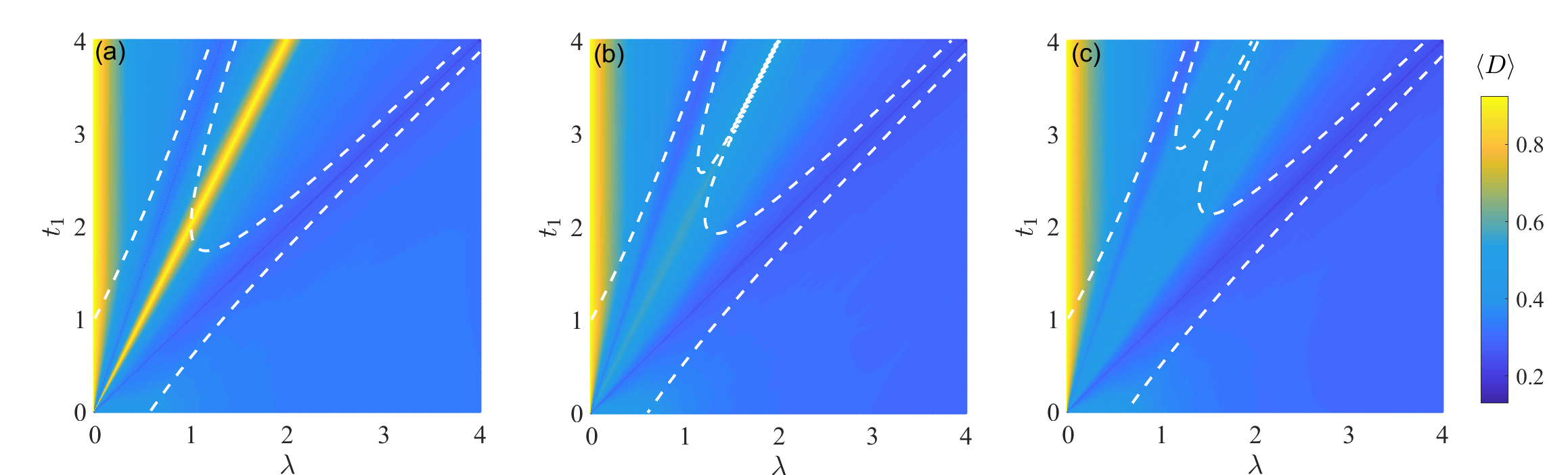}
	\caption{Mean fractal dimension $\langle D \rangle$ as a function of the intradimer hopping amplitude $t_1$ and disorder strength $\lambda$ for $p_1=0.5$, $\xi^{(1)}=\lambda$, $\xi^{(2)}=2\lambda$, and $\xi^{(3)}=3\lambda$. The white dashed lines denote the topological phase boundaries. The values of $p_2$ corresponding to panels (a)-(c) are $0$, $0.1$, and $0.2$, respectively.}
	\label{FigS6}
\end{figure*}

In this appendix, we analyze the localization properties of the generalized Bernoulli-disordered SSH model studied in the main text. In contrast to conventional disordered one-dimensional systems, where sufficiently strong disorder typically drives all eigenstates into exponentially localized Anderson states, the present model exhibits richer localization behavior, including reentrant extended regimes. To characterize these properties, we employ the fractal dimension.

For a large system, the fractal dimension of the $n$th eigenstate is defined as
\begin{equation}
	\label{EqS:fractal_dimension}
	D^{(n)}=-\lim_{N\rightarrow\infty}\frac{\ln\left(\textrm{IPR}^{(n)}\right)}{\ln(2N)}.
\end{equation}
Here, the inverse participation ratio is defined as
\begin{equation}
	\label{EqS:ipr_definition}	\textrm{IPR}^{(n)}=\sum_{i=1}^{N}\left(\left|\psi_{i,A}^{(n)}\right|^4+\left|\psi_{i,B}^{(n)}\right|^4\right),
\end{equation}
where $N$ is the total number of unit cells in the chain, and $\psi_{i,\sigma}^{(n)}$ denotes the amplitude of the $n$th eigenstate on sublattice $\sigma=A,B$ in the $i$th unit cell. From the scaling of $\textrm{IPR}^{(n)}$, one has $D^{(n)}\to 1$ for extended states and $D^{(n)}\to 0$ for localized states, while intermediate values $0<D^{(n)}<1$ indicate critical or intermediate behavior. The mean fractal dimension, averaged over the full spectrum, is defined as
\begin{equation}
	\label{EqS:mean_fractal_dimension}
	\langle D\rangle=\frac{1}{2N}\sum_{n=1}^{2N}D^{(n)}.
\end{equation}
Accordingly, $\langle D\rangle\to 1$ corresponds to an overall extended spectrum, $\langle D\rangle\to 0$ to a localized spectrum, and $0<\langle D\rangle<1$ to an intermediate regime.

Figure~\ref{FigS4}(a) shows the mean fractal dimension $\langle D\rangle$ as a function of the intradimer hopping amplitude $t_1$ and disorder strength $\lambda$ for a binary generalized Bernoulli distribution with $\xi^{(1)}=\lambda$, $\xi^{(2)}=2\lambda$, $p_1=2/5$, and $p_2=3/5$. The white dashed lines denote the topological phase boundaries. For a fixed $t_1$, as $\lambda$ increases, the system can evolve from an extended regime to a localized regime, then pass through a reentrant extended window, and finally enter a localized regime again. By comparing the localization pattern with the topological phase boundaries, one can clearly see that the localization behavior is not in one-to-one correspondence with the topological behavior. This demonstrates that the localization behavior in the present model is qualitatively richer than that in a conventional one-dimensional random system. To further examine the thermodynamic-limit behavior, Fig.~\ref{FigS4}(b) presents the finite-size scaling of the mean fractal dimension $\langle D\rangle$ for representative parameter sets. The dashed lines denote the extrapolation of $\langle D\rangle$ as a function of $1/\ln(2N)$, and the markers correspond to the same parameter sets indicated by the respective symbols in Fig.~\ref{FigS4}(a). The extrapolated values confirm the reentrant nature of the localization behavior inferred from Fig.~\ref{FigS4}(a). Figure~\ref{FigS4}(c) shows the localization phase diagram for another binary distribution, with the white dashed lines marking the topological phase boundaries. Similar to Fig.~\ref{FigS4}(a), the reentrant extended region identified from $\langle D\rangle$ does not coincide with the topological windows. This further confirms that the localization behavior and the topological behavior are not in one-to-one correspondence.

For the binary case, the reentrant extended regime is distributed around the relation
\begin{equation}
	\label{EqS:binary_extended_condition}
	t_1=\frac{\xi^{(1)}+\xi^{(2)}}{2}.
\end{equation}
This indicates that the position of the extended window is mainly controlled by the relative magnitudes of the two disorder values. To clarify the origin of this behavior, we consider the schematic configurations shown in Fig.~\ref{FigS5}. When $t_1=[\xi^{(1)}+\xi^{(2)}]/2$, the effective intradimer hoppings are redistributed into only two values, as illustrated in Fig.~\ref{FigS5}(b). In this regime, the chain can still be viewed as a generalized SSH model without strong effective bond breaking, so the eigenstates remain extended. This explains the robust extended window near the above condition. By contrast, when $t_1$ approaches $\xi^{(1)}$ or $\xi^{(2)}$, some effective intradimer hoppings become very small or vanish. The chain is then effectively fragmented into shorter segments, as illustrated in Fig.~\ref{FigS5}(c), which strongly suppresses wave-function propagation and enhances localization. Therefore, the reentrant localization behavior can be understood as arising from the competition between effective bond homogenization near $t_1=[\xi^{(1)}+\xi^{(2)}]/2$ and effective bond breaking near $t_1=\xi^{(1)}$ and $\xi^{(2)}$.

Figure~\ref{FigS6} shows $\langle D\rangle$ for a ternary generalized Bernoulli distribution with $p_1=0.5$, $\xi^{(1)}=\lambda$, $\xi^{(2)}=2\lambda$, and $\xi^{(3)}=3\lambda$, where the probability $p_2$ is varied. The white dashed lines again denote the topological phase boundaries. When $p_2=0$, the model reduces to a binary disorder distribution and exhibits a clear reentrant extended window, as shown in Fig.~\ref{FigS6}(a). As $p_2$ increases, the localization tendency is enhanced and the extended window is gradually suppressed, as seen in Figs.~\ref{FigS6}(b) and \ref{FigS6}(c). Comparing the localization pattern with the topological phase boundaries again shows that the localization behavior does not coincide with the topological one. These results demonstrate that the emergence of reentrant extended regimes is highly sensitive to the detailed structure of the multivalued disorder distribution, but it is not directly tied to the locations of the topological phase boundaries.

\begin{figure*}[htbp]
	\centering
	\includegraphics[width=12cm]{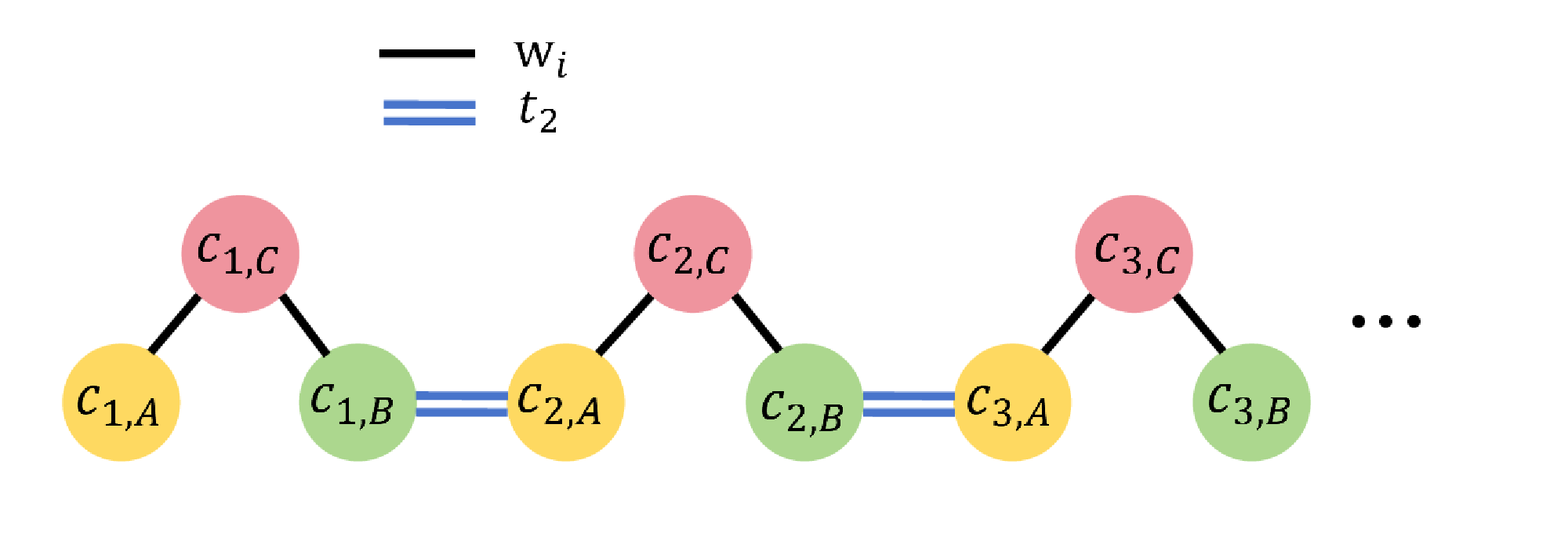}
	\caption{Schematic illustration of the proposed photonic-waveguide implementation. Each unit cell contains three waveguides labeled $A$, $B$, and $C$, where $C$ is an auxiliary waveguide. The coupling between $C$ and $A$ or $B$ in the $i$th unit cell is denoted by $\textrm{w}_i$, while the coupling between neighboring unit cells is the fixed interdimer coupling $t_2$.}
	\label{FigS7}
\end{figure*}

\section{POSSIBLE PHOTONIC IMPLEMENTATION}

The SSH model and its variants have been experimentally realized in a variety of platforms, including cold-atom systems~\cite{MeierEJ2018}, photonic and acoustic lattices~\cite{HuB2023,YuZ2026S}, and electric circuits~\cite{YangH2022S}. Here we outline a possible photonic-waveguide implementation of the generalized Bernoulli-disordered SSH model studied in the main text. This discussion is intended as a feasible route for realizing the effective Hamiltonian and probing its dynamical signatures; the topological results presented above do not rely on the details of this implementation. As illustrated in Fig.~\ref{FigS7}, each unit cell contains three waveguides, labeled $A$, $B$, and $C$. The waveguides $A$ and $B$ form the effective SSH dimer, while the auxiliary waveguide $C$ is introduced to engineer the effective intradimer coupling.

We denote by $\textrm{w}_i$ the coupling between the auxiliary waveguide $C$ and waveguide $A$ or $B$ in the $i$th unit cell. The coupling between waveguides belonging to neighboring unit cells is taken to be a fixed interdimer coupling $t_2$. The propagation-constant detunings of the three waveguides are denoted by $\Delta_{i,A}$, $\Delta_{i,B}$, and $\Delta_{i,C}$. Under the coupled-mode description, the light dynamics in the waveguide array is governed by
\begin{equation}
	\label{EqS:waveguide_full_model}
	\begin{aligned}
		-i\frac{dc_{i,A}}{dz} & = \Delta_{i,A}c_{i,A}+t_2c_{i-1,B}+\textrm{w}_i c_{i,C}, \\
		-i\frac{dc_{i,B}}{dz} & = \Delta_{i,B}c_{i,B}+t_2c_{i+1,A}+\textrm{w}_i c_{i,C}, \\
		-i\frac{dc_{i,C}}{dz} & = \Delta_{i,C}c_{i,C}+\textrm{w}_i\left(c_{i,A}+c_{i,B}\right),
	\end{aligned}
\end{equation}
where $c_{i,\sigma}$ denotes the field amplitude in waveguide $\sigma=A,B,C$ of the $i$th unit cell, and $z$ is the propagation distance.

When the detuning of the auxiliary waveguide is much larger than its coupling strength, namely $|\Delta_{i,C}|\gg \textrm{w}_i$, the $C$ mode can be adiabatically eliminated. In this limit, the three-waveguide model reduces to an effective two-sublattice model described by
\begin{equation}
	\label{EqS:waveguide_effective_model}
	\begin{aligned}
		-i\frac{dc_{i,A}}{dz} & =\left(\Delta_{i,A}-\frac{\textrm{w}_i^2}{\Delta_{i,C}}\right)c_{i,A}
		-\frac{\textrm{w}_i^2}{\Delta_{i,C}}c_{i,B}
		+t_2c_{i-1,B}, \\
		-i\frac{dc_{i,B}}{dz} & =\left(\Delta_{i,B}-\frac{\textrm{w}_i^2}{\Delta_{i,C}}\right)c_{i,B}
		-\frac{\textrm{w}_i^2}{\Delta_{i,C}}c_{i,A}
		+t_2c_{i+1,A}.
	\end{aligned}
\end{equation}

To reproduce the model considered in the main text, the parameters can be chosen such that
\begin{equation}
	\label{EqS:waveguide_parameter_choice}
	\Delta_{i,A}=\Delta_{i,B}=\frac{\textrm{w}_i^2}{\Delta_{i,C}}=-t_1+\xi_i.
\end{equation}
Under this condition, the on-site terms in Eq.~(\ref{EqS:waveguide_effective_model}) vanish, and one obtains
\begin{equation}
	\label{EqS:waveguide_target_model}
	\begin{aligned}
		-i\frac{dc_{i,A}}{dz} & = \left(t_1-\xi_i\right)c_{i,B}+t_2c_{i-1,B}, \\
		-i\frac{dc_{i,B}}{dz} & = \left(t_1-\xi_i\right)c_{i,A}+t_2c_{i+1,A},
	\end{aligned}
\end{equation}
which is exactly the coupled-mode form of the SSH model with disordered intradimer hopping amplitudes studied in the main text.

Therefore, the effective intradimer hopping can be engineered through the detunings and the couplings to the auxiliary waveguides, while the interdimer hopping is controlled by the spacing between neighboring waveguides. This provides a feasible photonic platform for implementing the generalized Bernoulli-disordered SSH model and probing its reentrant topological behavior experimentally.

\end{document}